\documentclass[12pt,preprint]{aastex}
\usepackage{natbib}
\bibpunct[;]{(}{)}{;}{a}{}{,} 

\begin{document}

\def\HII{\ion{H}{2}}
\setcounter{table}{0}

\title{The Wyoming Survey for H$\alpha$.  III.  A Multi-wavelength Look at Attenuation by Dust in Galaxies out to $z \sim 0.4$}
\author{Carolynn~A.~Moore\altaffilmark{1, 2}, Daniel~A.~Dale\altaffilmark{1}, Rebecca~J.~Barlow\altaffilmark{1}, Seth~A.~Cohen\altaffilmark{1, 3}, David~O.~Cook\altaffilmark{1}, L.~C.~Johnson\altaffilmark{1, 4}, ShiAnne~M.~Kattner\altaffilmark{1}, Janice~C.~Lee\altaffilmark{5}, Shawn~M.~Staudaher\altaffilmark{1}}
\altaffiltext{1}{Dept. of Physics \& Astronomy, University of Wyoming, Laramie, WY 82071}
\altaffiltext{2}{e-mail: garcia@uwyo.edu}
\altaffiltext{3}{current address: Dept. of Physics \& Astronomy, Dartmouth College, Hanover, NH 03747}
\altaffiltext{4}{current address: Dept. of Astronomy, University of Washington, Seattle, WA 98195}
\altaffiltext{5}{Hubble Postdoctoral Research Fellow, Carnegie Observatories, Pasadena, CA 91101}

\begin{abstract}
We report results from the Wyoming Survey for H$\alpha$ (WySH), a comprehensive four-square degree survey to probe the evolution of star-forming galaxies over the latter half of the age of the Universe.  We have supplemented the H$\alpha$ data from WySH with infrared data from the {\it Spitzer} Wide-area Infrared Extragalactic (SWIRE) Survey and ultraviolet data from the Galaxy Evolution Explorer (GALEX) Deep Imaging Survey.  This dataset provides a multi-wavelength look at the evolution of the attenuation by dust, and here we compare a traditional measure of dust attenuation ($L(TIR)/L(FUV)$) to a diagnostic based on a recently-developed robust star formation rate (SFR) indicator, $[\rm H\alpha_{obs}+24\mu{m}]/H\alpha_{obs}$.  With such data over multiple epochs, the evolution in the attenuation by dust with redshift can be assessed.  We present results from the ELAIS-N1 and Lockman Hole regions at $z~\sim~0.16$, $0.24$, $0.32$ and $0.40$.  While the ensemble averages of both diagnostics are relatively constant from epoch to epoch, each epoch individually exhibits a larger attenuation by dust for higher star formation rates.  Hence, an epoch to epoch comparison at a fixed star formation rate suggests a mild decrease in dust attenuation with redshift.  
\end{abstract}



\section{INTRODUCTION}
\label{intro}
A fundamental property of a galaxy is the number of stars it gives birth to each year, commonly known as its star formation rate (SFR).  Likewise, the average number of stars forming per year per cubic megaparsec, the cosmic star formation rate density (SFRD), is a key parameter in the study of galaxy evolution.  To quantify these fundamental parameters, a number of different star formation rate indicators are used, and they stem from a wide range of wavelengths.  The star formation rate can be estimated using the X-ray, ultraviolet (UV), infrared (IR) or radio continuua, a variety of Hydrogen recombination lines, or heavier-atomic forbidden lines \citep[e.g.,][and references therein]{Kennicutt:1998, Cardiel:2003, Iglesias-Paramo:2004, Ly:2007}.  Three of the most widely-used star-forming measures are the ultraviolet and infrared continua and the H$\alpha$ recombination line, which respectively probe i) the emission from hot, relatively young (OB) stars, ii) starlight that has been reprocessed by interstellar dust and iii) the \HII~regions.  Because these three star formation rate indicators arise from three different processes and emit over three different wavelength regimes, there are known discrepancies between them \citep{Hopkins:2001a,Tresse:2002, Teplitz:2003}.  The observed star formation rates will clearly disagree due to the variable effect attenuation by dust has as a function of wavelength.  But even after proper extinction corrections are made, additional factors may lead to differences in the star formation rate, including old stellar populations contributing to the infrared emission, incorrect Initial Mass Function (IMF) assumptions, differences in the relative stellar/nebular/dust geometries, contamination from Active Galactic Nuclei (AGN), and the differential impacts of metallicity and the ionization state of the gas on the various SFR indicators \citep{Kennicutt:1998, Xu:2003, Choi:2006, Salim:2007, Lara-Lopez:2009}.  While many of these discrepancies are due to the uncertainties in the conversion factors from luminosities to star formation rates, an additional possibility for discrepancy exists within the different survey selection criteria \citep{Tresse:2002, Buat:2007b}.  More `robust' star formation rate measures have recently been devised using a combination of infrared~+~ultraviolet data or infrared~+~H$\alpha$ data \citep[][and references therein]{Hirashita:2003, Calzetti:2007, Kennicutt:2007, Prescott:2007, Kennicutt:2009} because the infrared emission traces extinguished starlight while the ultraviolet and H$\alpha$ emission trace relatively unextinguished starlight.  

These SFR indicators have been used in a large number of surveys to map the cosmic SFR at a variety of redshifts \citep[e.g.,][]{Cowie:2004, LeFloch:2005, Doherty:2006, Dale:2008, Prescott:2009}.  The cumulative result of these efforts show that the cosmic SFRD has significantly evolved over the last several gigayears, perhaps by as much as a factor of 10 between $z~\sim~1$ and the present epoch \citep[e.g.,][]{Watson:2009}.  Therefore, it is not surprising that the galaxies we observe locally do not necessarily resemble those at higher redshifts \citep{Pei:1999, Davies:2009}.  For example, Luminous Infrared Galaxies (LIRGs) and Ultra-Luminous Infrared Galaxies (ULIRGs) are increasingly important as we look back to redshifts of $z~\sim~1-2$ as both the IR and UV luminosity functions evolve toward a predominance of LIRGs at higher $z$ \citep{Papovich:2006, Reddy:2006, Reddy:2008, Buat:2009, Salim:2009}.  These dusty objects come to dominate the luminosity of the galaxy population as we move toward higher redshifts.  Does this effect lead to a change in the average amount of obscuration by dust over time?  Does the build up of metal content within galaxies as they age strongly affect their extinction by dust \citep[e.g.,][]{Mannucci:2009}?  Not only do galaxies evolve with redshift, they each evolve on their own timescales (i.e., they run out of gas and cease to form stars, they are involved in mergers with other galaxies, etc.).  These changes imply that the light that is detected from a given galaxy will be a function not only of its redshift, but also a function of parameters such as metallicity, star formation history, merger stage (if applicable), and the relative luminosity dominance of any supermassive black hole \citep[][and references therein]{Kennicutt:1998, Calzetti:1999, Pei:1999, Jansen:2001, Dale:2007, Wu:2007, Engelbracht:2008}.

Not only do SFR indicators tell us how many stars are being formed, but they can provide other physical insight when two different SFR indicators are coupled together.  For example, the infrared-to-ultraviolet ratio is typically used to measure the amount of dust attenuation at ultraviolet wavelengths.  Studies coupling infrared and ultraviolet data have shown that the slope of the ultraviolet continuum is a useful probe of the attenuation by dust in starburst galaxies \citep{Calzetti:1994, Meurer:1999}, so simply tracking the infrared-to-ultraviolet ratio with lookback time should prove fruitful.  Recent efforts, however, lead to a confusing set of divergent results.  Evidence from (Ôrest frameÕ) ultraviolet-selected surveys by \citet{Buat:2007b}, \citet{Burgarella:2007}, and \citet{Reddy:2006} at $z~\sim~0, 1,$ and 2, respectively, shows that the average dust attenuation within galaxies decreases with increasing redshift.  Similar results were reported by \citet{Buat:2007a} and \citet{Xu:2007} for LIRGs.  This is the expected result if the dust within galaxies slowly builds up as the effects of generations of stellar lifecycles accumulate over time.  However, \citet{LeFloch:2005} find that the comoving infrared energy density for an infrared-selected sample evolves as $(1+z)^{3.9}$ at $0<z<1$, whereas the ultraviolet only evolves as $(1+z)^{2.5}$ over the same span in redshift \citep{Schiminovich:2005}.  \citet{Takeuchi:2005}, using the same dataset as \citet{LeFloch:2005}, report a similar result over the same range in redshift.  While the \citet{LeFloch:2005} and \citet{Schiminovich:2005} results are based on samples of galaxies that are likely very different, together, they could imply that the average ultraviolet attenuation by dust within galaxies increases with lookback time.  This result would be counterintuitive if dust attenuation were only governed by metal abundance.  However, the result is understandable if the attenuation by dust increases with SFR \citep[e.g.,][]{Wang:1996, Heckman:1998, Calzetti:2001, Hopkins:2001b} and the cosmic SFRD increases with lookback time.  Finally, a third possible result is reported by \citet{Bell:2005}, \citet{Choi:2006}, \citet{Xu:2007}, and \citet{Zheng:2007} who find dust attenuation values at higher redshifts ($0.6<z<0.8$) consistent with those found in the current epoch.  However, it should be noted that all of these studies utilize some sort of IR selection.  While not all of the studies using an IR selection come to this same conclusion, this possible selection effect should be looked at in more depth. 

\citet{Buat:2007b} have shown that ultraviolet and infrared surveys of the nearby universe do not necessarily sample the same galaxy populations, and thus the above discrepancies may be in part due to sample selection.  The ultraviolet continuum traces the star formation rate on a time scale of $\sim$$10^{8}$ years \citep[e.g.,][]{Iglesias-Paramo:2004}, while the infrared continuum traces star formation rates on a time scale of $\sim$$10^{7}$ years \citep[e.g.,][]{Kennicutt:1998}.  This, as well as issues such as feedback from star formation rate evolution, Lyman $\alpha$ absorption, the evolution of dust grains and the evolution of the average dust temperature can all affect the infrared and ultraviolet luminosities.  Invoking data from a large, uniform H$\alpha$-selected survey could help to clear this confusion \citep{Iglesias-Paramo:2004}.  H$\alpha$ emission traces star formation rates on a time scale of $\lesssim$$10^{7}$ years \citep[e.g.,][]{Iglesias-Paramo:2004}, making it an ideal comparison to the infrared continuum.  Focusing just on this similarity in timescales, and ignoring evolutionary processes such as the dispersal of dust grains into the ISM, the ratio of H$\alpha$ to IR should remain relatively constant over time within a galaxy.  Recent work from the {\it Spitzer} Infrared Nearby Galaxy Survey (SINGS) find that the H$\alpha$+24$\mu$m data have an almost one-to-one correlation with that of extinction corrected Paschen~$\alpha$, considered an ÒidealÓ star formation rate indicator as it directly traces \HII~regions and lies at a relatively long wavelength to help minimize the effects of obscuration by dust \citep[][see Fig. 11 in Calzetti et al. 2007]{Calzetti:2007, Kennicutt:2007, Prescott:2007}.  This H$\alpha$+24$\mu$m-based star formation rate indicator can be compared to the more commonly used indicator, $L({\rm H}\alpha)$ alone \citep{Kennicutt:1998} to form a diagnostic sensitive to the amount of attenuation by dust within galaxies, a multi-wavelength diagnostic where all the inputs conveniently trace star formation on the same time scale.  Although recent results from \citet{Kennicutt:2009} indicate that it can be just as effective to use a combination of the total infrared (TIR) emission~+~H$\alpha$ or 8$\mu$m~+~H$\alpha$, TIR data are not frequently available for surveys of distant galaxies and 8$\mu$m data may be more problematic due to the variable nature of polycyclic aromatic hydrocarbon (PAH) emission with radiation field hardness and metallicity \citep{Engelbracht:2005, Madden:2006, OHalloran:2006, Rosenberg:2006, Wu:2006, Dale:2009a}    

The Wyoming Survey for H$\alpha$ (WySH) is a large-area, ground-based, narrowband optical imaging survey for H$\alpha$ at redshifts $z~\sim~0.16$, $0.24$, $0.32$, and $0.40$ \citep{Dale:2008}.  In addition, this effort is being supplemented by the NEWFIRM Narrowband H$\alpha$ Survey, a collaborative near-infrared imaging survey of H$\alpha$ at redshifts of $z~\sim~0.81$ and $2.2$ \citep[][Lee et al., in prep; Ly et al., in prep.]{Lee:2009a}.  Both surveys target key extragalactic deep fields where an abundance of multiwavelength imaging and spectroscopy are already available.  In this work we utilize $(\rm H\alpha_{obs}+24\mu m)/H\alpha_{obs}$ to directly gauge the amount of attenuation by dust in a large sample of galaxies that span redshifts out to $z~\sim~0.4$.  Our main goal is to robustly measure any changes in the average dust attenuation over cosmic time.  Secondary goals include comparing the relatively new technique for assessing the attenuation by dust, by combining H$\alpha$ and 24$\mu$m, with schemes based on the coupling of IR and UV.  

In this paper, we outline the datasets used in \S\ref{data}, present the analysis and initial results from $z~\sim~0.16$, $0.24$, $0.32$ and $0.40$ in \S\ref{anal-res}, and summarize our findings in \S\ref{sum}.  The cosmology used throughout this paper is $H_{0}=70$ $h_{70}$ km s$^{-1}$ Mpc$^{-1}$, $\Omega_{m}=0.3$, and $\Omega_{\lambda}=0.7$.
 

\section{DATA}
\label{data}
\subsection{Fields}
\begin{figure}
\includegraphics[scale=0.64]{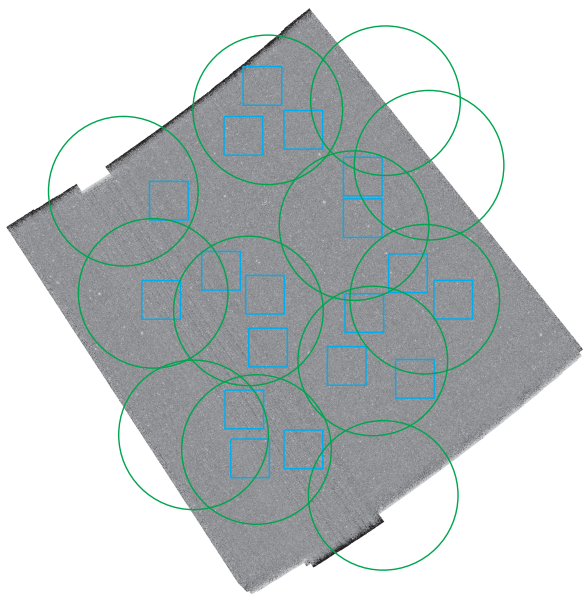}
\caption{The 24$\mu$m image from SWIRE of the ELAIS-N1 region ($\alpha \sim 16^{\rm h}11^{\rm m}$, $\delta \sim +55^{\circ}00\arcmin\ $) overlayed with circles to represent GALEX fields and squares to indicate the WySH fields.  North is up and east is to the left.  Dimensions of the SWIRE image are approximately $3^{\circ}$ by $3^{\circ}$, the GALEX overlays have diameters of approximately $1^{\circ}$, and the WySH overlays have dimensions of $18\arcmin\ $by $18\arcmin\ $.}
\label{elaisoverlay}
\end{figure}

\begin{figure}
\includegraphics[scale=0.64]{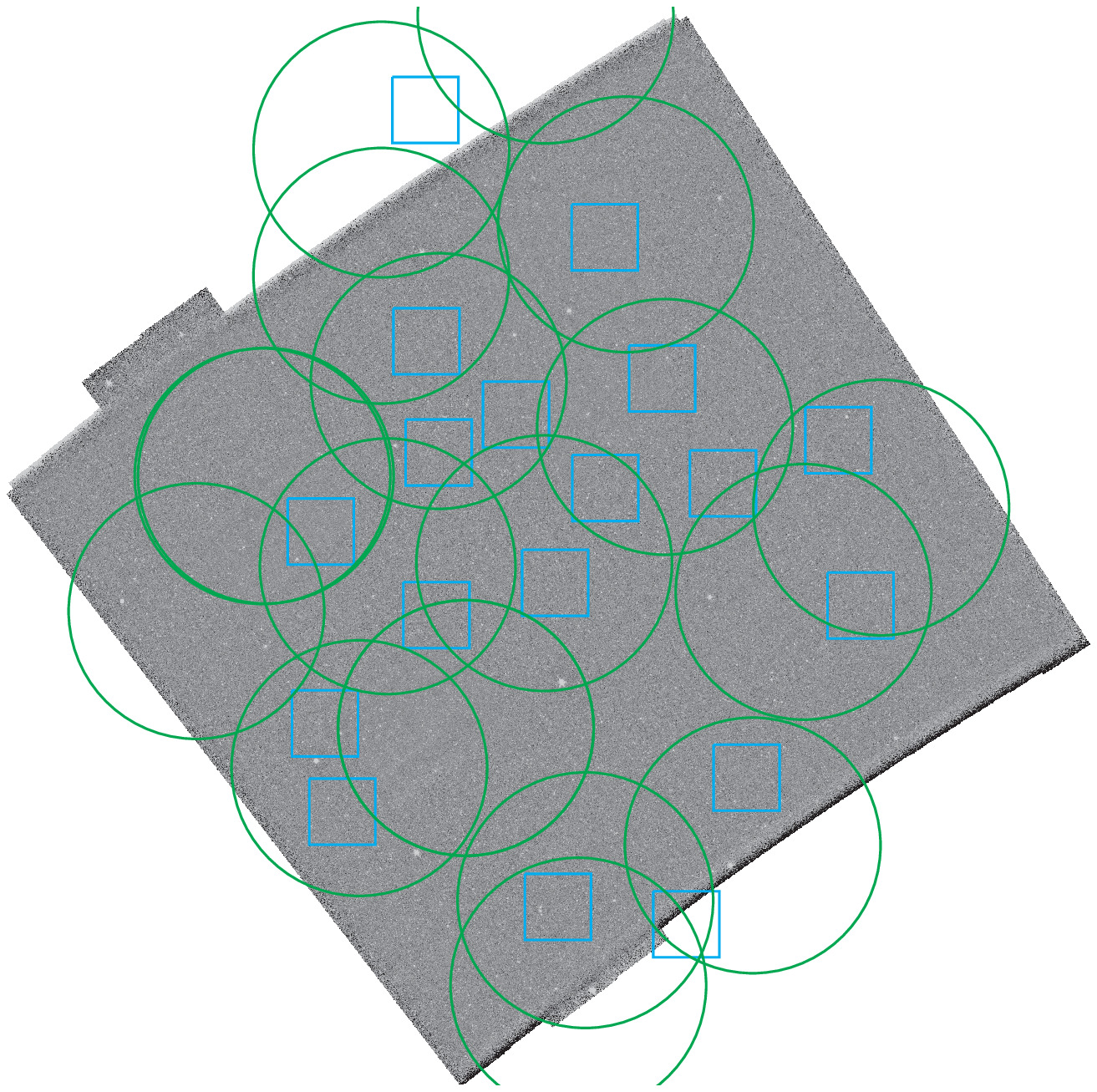}
\caption{The 24$\mu$m image from SWIRE of the Lockman Hole region ($\alpha \sim 10^{\rm h}45^{\rm m}$, $\delta \sim +58^{\circ}00\arcmin\ $) overlayed with circles to represent GALEX fields and squares to indicate the WySH fields.  North is up and east is to the left.  Dimensions of the SWIRE image are approximately $3.6^{\circ}$ by $3^{\circ}$, the GALEX overlays have diameters of approximately $1^{\circ}$, and the WySH overlays have dimensions of $18\arcmin\ $by $18\arcmin\ $.}
\label{lockmanoverlay}
\end{figure}

Results for the ELAIS-N1 ($\alpha \sim 16^{\rm h}11^{\rm m}$, $\delta \sim +55^{\circ}00\arcmin\ $) and Lockman Hole ($\alpha \sim 10^{\rm h}45^{\rm m}$, $\delta \sim +58^{\circ}00\arcmin\ $) regions are presented here.  These regions are popular targets for deep extragalactic surveys since they are comparatively free of contamination by zodiacal and Galactic dust ($I_{\nu}$(100$\mu$m)$\sim$1~MJy~sr$^{-1}$), as well as by bright stars and nearby galaxies.  Two relevant surveys carried out in ELAIS-N1 and the Lockman Hole are the {\it Spitzer} Wide-Area Infrared Extragalactic (SWIRE) Survey (see \S~\ref{ir}) and the Galaxy Evolution Explorer (GALEX) Deep Imaging Survey (see \S~\ref{uv}).  Some thirty separate $18\arcmin\ $fields were targeted by the WySH program in these two fields (see Figures~\ref{elaisoverlay} and \ref{lockmanoverlay}).  Table~\ref{wyshsum} summarizes the total coverage at each epoch for both fields.

\begin{deluxetable}{lcccccc}
\rotate
\tabletypesize{\scriptsize}
\tablecaption{WySH Optical Imaging \& Source Counts.}
\tablehead{
\colhead{$z$} &
\colhead{ELAIS-N1 ($\square^{\circ}$)} &
\colhead{Lockman Hole ($\square^{\circ}$)} &
\colhead{NB int. time (s)} &
\colhead{\# H$\alpha$\tablenotemark{a}} &
\colhead{\# H$\alpha$\tablenotemark{a} + 24$\mu$m\tablenotemark{b}} &
\colhead{\# H$\alpha$\tablenotemark{a} + 24$\mu$m\tablenotemark{b} + UV\tablenotemark{c}}
}
\startdata
0.16 &1.38 &1.01 &1200 &94 &43 &39\\
0.24 &1.35 &0.83 &3600 &239 &69 &62\\
0.32 &1.33 &0.87 &6000 &250 &82 &67\\
0.40 &0.52 &0.00 &9600 &49 &20 &16\\
\hline
{\it Totals}:&-- &-- &-- &632 &214 &184\\
\enddata
\tablenotetext{a}{Detections from WySH are above 3$\sigma$.}
\tablenotetext{b}{Detections from SWIRE are above 3$\sigma$.}
\tablenotetext{c}{Detections from GALEX are above 5$\sigma$.}
\label{wyshsum}
\end{deluxetable}

\subsection{Datasets}
\subsubsection{H$\alpha$} 
WySH is a $\sim$4 square degree, ground-based, narrowband imaging survey for H$\alpha$ emitting galaxies over the latter half of the age of the universe.  WySH uses a pair of wavelength-adjacent narrowband filters at each redshift epoch (for $z~\lesssim~0.40$) for improved continuum subtraction, relative to using a broadband filter to subtract the continuum.  The epochs being observed at the Wyoming Infrared Observatory (WIRO) 2.3m telescope are $z \sim 0.16, 0.24, 0.32,$ and 0.40 \citep{Dale:2008}.  This optical survey is being supplemented with collaborative near-infrared work at the Kitt Peak National Observatory 4m telescope at $z \sim 0.81$ and $2.2$ by the NEWFIRM Narrowband H$\alpha$ Survey\footnote{http://newfirm.ociw.edu/wiki/Main\_Page}, which is observing through low airglow windows at 1.19 and 2.10$\mu$m (Lee et al., in prep.; Ly et al., in prep.).  These data are being used to constrain the cosmic star formation rate density out to $z~\sim~2.2$.  WySH data are sensitive down to a 3$\sigma$ survey depth of $\lesssim$1${\rm M_{\odot}}$~yr$^{-1}$ for $z~\sim~0.16, 0.24, 0.32$ and $0.40$, using the \citet{Kennicutt:1998} equation to convert from $L({\rm H}\alpha)$ to SFR(H$\alpha$).  This value has not been corrected for dust attenuation.  For a full description of the optical survey and preliminary results for the SFRD and luminosity function at $z~\sim~0.16, 0.24, 0.32$ and $0.40$, see \citet{Dale:2008, Dale:2010}.

\subsubsection{Infrared}
\label{ir} 
SWIRE \citep{Lonsdale:2003}, one of the {\it Spitzer} Legacy Projects, was designed to image $\sim$49 square degrees of the sky, as well as detect very distant galaxies (out to $z~\sim~3$) in order to look at different properties of galaxy (and star) formation and evolution\footnote{www.ipac.caltech.edu/SWIRE/}.  SWIRE uses all seven imaging bands on {\it Spitzer} (3.6, 4.5, 5.8 and 8.0$\mu$m with IRAC and 24, 70 and 160$\mu$m with MIPS).  Using the \citet{Calzetti:2007} relation between $L_{\nu}(24\mu m)$ and SFR,
\begin{equation}
  SFR ({\rm M_{\odot}~yr^{-1}}) = 1.27 \times 10^{-38}[\nu L_{\nu}(24\mu{\rm m}) ({\rm erg~s^{-1}})]^{0.8850}, \\
\label{24_sfr}
\end{equation}  
we find that, for a 3$\sigma$ 24$\mu$m survey depth of 135$\mu$Jy, at $z~\sim~0.16, 0.24, 0.32$ and $0.40$, galaxies with SFRs of approximately 0.22, 0.50, 0.86 and 1.4~${\rm M_{\odot}}$~yr$^{-1}$ (respectively) can be detected.  The SWIRE survey covers 9.2 square degrees in the ELAIS-N1 field and 11 square degrees in the Lockman Hole field (see Figures \ref{elaisoverlay} and \ref{lockmanoverlay}).

\subsubsection{Ultraviolet}
\label{uv}
GALEX \citep{Martin:2005} is a space-based mission designed to study the ultraviolet properties of galaxies in the local universe\footnote{www.galex.caltech.edu}.  Utilizing both the far- and near-ultraviolet (FUV, NUV) detectors (1538.6\AA~and 2315.7\AA, and FWHMs of 269\AA~and 616\AA~respectively), the Deep Imaging Survey pointings have exposure times of 30ks, as opposed to the 0.1ks exposure times per pointing for the GALEX All Sky Survey.  Deep Imaging Survey pointings include ELAIS-N1 and the Lockman Hole.  The Deep Imaging Survey has 5$\sigma$ limiting magnitudes (in $m_{AB}$) of $m_{NUV}$=24.4 and $m_{FUV}$=24.8, angular resolutions of 5.3\arcsec\ (NUV) and 4.2\arcsec\ (FUV) and samples galaxies with an average redshift of $<z>=0.85$ \citep{Martin:2005, Morrissey:2007}.  Using the FUV limiting magnitude and the \citet{Kennicutt:1998} relation between UV luminosity and SFR, we find that, for a 5$\sigma$ survey depth of $F_{\nu}=4.365~\times~10^{-33}~{\rm W~m^{-2}~Hz^{-1}}$, at $z~\sim~0.16, 0.24, 0.32$ and $0.40$, SFRs of approximately 0.04, 0.11, 0.20 and 0.35~${\rm M_{\odot}}$~yr$^{-1}$ (respectively) can be detected (assuming no attenuation by dust).  It is also important to note that the GALEX Deep Imaging Survey is confusion limited and that standard pipeline results are not reliable for faint ($m_{NUV} > 23$~mag) sources \citep{Burgarella:2007, Martin:2007, Zamojski:2007}.  This may slightly underestimate dust attenuation using the $\nu L_{\nu}(24\mu{\rm m})/\nu L_{\nu}(FUV)$ indicator for the UV faint galaxies.  These UV faint galaxies follow the same trend as the brighter ($m_{NUV} < 23$~mag) UV galaxies, and make up $\sim$8\% of the entire sample ([0, 5, 8, 2] sources at $z\sim$[0.16, 0.24, 0.32, 0.40]).  For the joint WySH/GALEX Deep Imaging Survey analysis carried out here, a total of 12 $\sim$1 degree fields covering the ELAIS-N1 region and 18 $\sim$1 degree fields covering the Lockman Hole region (see Figures \ref{elaisoverlay} and \ref{lockmanoverlay}) are used.

\subsubsection{Merged Catalog}
The H$\alpha$ data from WySH were used for the initial sample selection, providing a sample of 94~+~239~+~250~+~49 galaxies at $z \sim 0.16, 0.24, 0.32,$ and 0.40, respectively based on a 3$\sigma$ detection limit.  This data was first merged with a separate 24$\mu$m SWIRE catalog with a depth of 3$\sigma$. Then, using the GATOR tool\footnote{http://irsa.ipac.caltech.edu/applications/Gator/} (provided through the NASA/IPAC Infrared Science Archive (IRSA)), the Spring 2005 SWIRE {\it Spitzer} catalogs for both the ELAIS-N1 and the Lockman Hole fields were accessed.  These catalogs contain data in the four IRAC bands, as well as the MIPS 24$\mu$m band.  All sources in the catalog contain 3.6$\mu$m and 4.5$\mu$m data, but not all sources were detected at the necessary level in the remaining bands.  In GATOR, we uploaded the WySH source coordinates separately for each epoch ($z~\sim~0.16, 0.24, 0.32, 0.40$), and searched for coordinate matches with SWIRE sources within a 3\arcsec\ radius to create a WySH~+~SWIRE database.  To create the WySH~+~SWIRE~+~GALEX merged catalog of sources,  the WySH~+~SWIRE catalogs were coordinate matched with the GALEX catalogs from GALEX Release 5 (GR5) using a search radius of 3\arcsec .  The distance between any two H$\alpha$ sources is larger than the point-spread functions (PSFs) of both {\it Spitzer} and GALEX, so the $\sim$6\arcsec\ resolution of MIPS24 and GALEX imaging usually does not result in multiple galaxies contributing to a single IR or UV flux.  However, if multiple SWIRE or GALEX counterparts are found, the closest one is used.  For H$\alpha$ sources having H$\alpha$ neighbors within 3\arcsec\ (7 sources), there are 3 UV sources which appear as extended (blended) sources, which accounts for $< 2\%$ of the entire sample.  The sample is further reduced in size by requiring sources to have data in both UV bands as well as the 24$\mu$m band.  Both the WySH~+~SWIRE and WySH~+~SWIRE~+~GALEX catalogs are used in this research where appropriate.  Ultimately, this can be considered a 24$\mu$m selection.  The sample sizes are listed in Table~\ref{wyshsum}. 

\subsection{Corrections}
\subsubsection{{\it k} Corrections}
\begin{figure}
\includegraphics{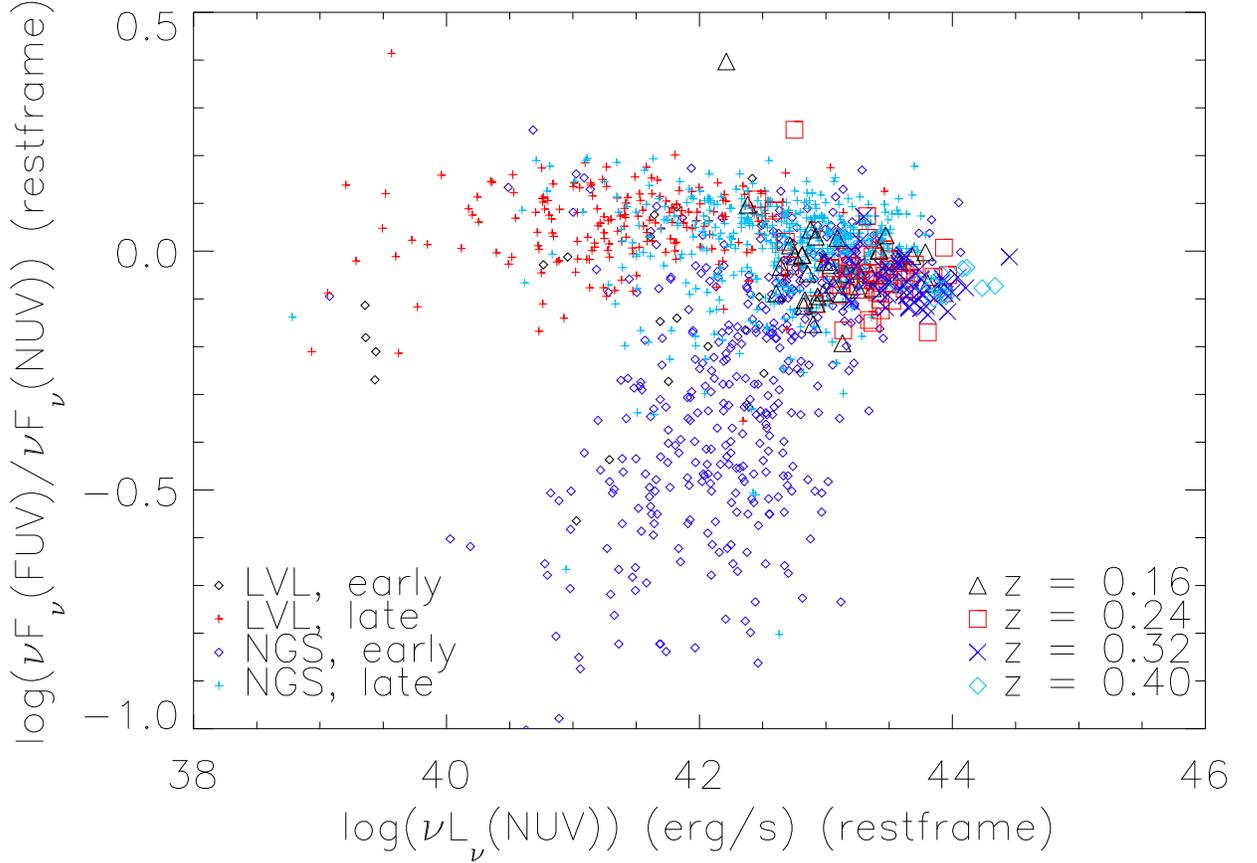}
\caption{The ratio of $\log (\nu F_{\nu}(FUV)/\nu F_{\nu}(NUV))$ v. $\log (\nu L_{\nu}(NUV))$.  Small pluses represent data from late type (T-type $\geq$ 3) galaxies in the {\it Spitzer} Local Volume Legacy survey and NGS, while small diamonds represent early type (T-type $<$ 3) galaxies in the {\it Spitzer} Local Volume Legacy survey and NGS.  These local data show that $\log (\nu F_{\nu}(FUV)/\nu F_{\nu}(NUV))$ remains relatively flat with median values (for the late type galaxies) of 0.06 dex for the {\it Spitzer} Local Volume Legacy survey and 0.02 dex for NGS, with respect to $\log (\nu L_{\nu}(NUV))$.  Rest-frame data from WySH are also shown.  Triangles represent data at $z\sim0.16$, squares represent data at $z\sim0.24$, crosses represent data at $z\sim0.32$ and diamonds represent data at $z\sim0.40$.}
\label{uvcorrect}
\end{figure}

For {\it k} corrections, the central redshift value for each epoch is used rather than the redshift for each source within a given epoch since $\delta z$ per epoch is $\sim$0.02 and the redshifts for this sample have been determined photometrically \citep{Rowan-Robinson:2008}.  The ultraviolet data for the sample were {\it k} corrected by calculating the slope between the observed FUV and NUV fluxes and by extrapolating from that slope to the rest frame FUV and NUV fluxes.  The FUV {\it k} corrections are modest and have median values of $f_{\nu}(FUV)^{true}/f_{\nu}(FUV)^{obs}=[0.91, 0.89, 0.86, 0.85]$ at $z~\sim$~[0.16, 0.24, 0.32, 0.40].  Comparing this to local data from the {\it Spitzer} Local Volume Legacy (LVL) survey \citep[][Lee et al. 2009, in prep.; for H$\alpha$, 24$\mu$m and UV data, respectively]{Kennicutt:2008, Dale:2009b}, and the Nearby Galaxy Survey \citep[NGS;][]{GildePaz:2007} in Figure~\ref{uvcorrect}, the ultraviolet continuum of late-type galaxies (T-type~$\geq$~3) is also relatively flat (median values along the y-axis are 0.06 dex for the {\it Spitzer} Local Volume Legacy survey and 0.02 dex for NGS).  (Also see Figure 2 of \citet{Dale:2007} and Figures 5 and 6 of \citet{GildePaz:2007}.)  For the infrared data, the {\it k} corrections are derived based on the average star-forming SED models of \citet{Dale:2002}.  The 24$\mu$m {\it k} corrections are $f_{\nu}(24)^{true}/f_{\nu}(24)^{obs}=[1.57, 1.90, 2.25, 2.63]$ at $z~\sim$~[0.16, 0.24, 0.32, 0.40].

\subsubsection{AGN Corrections}
\begin{figure}
\includegraphics{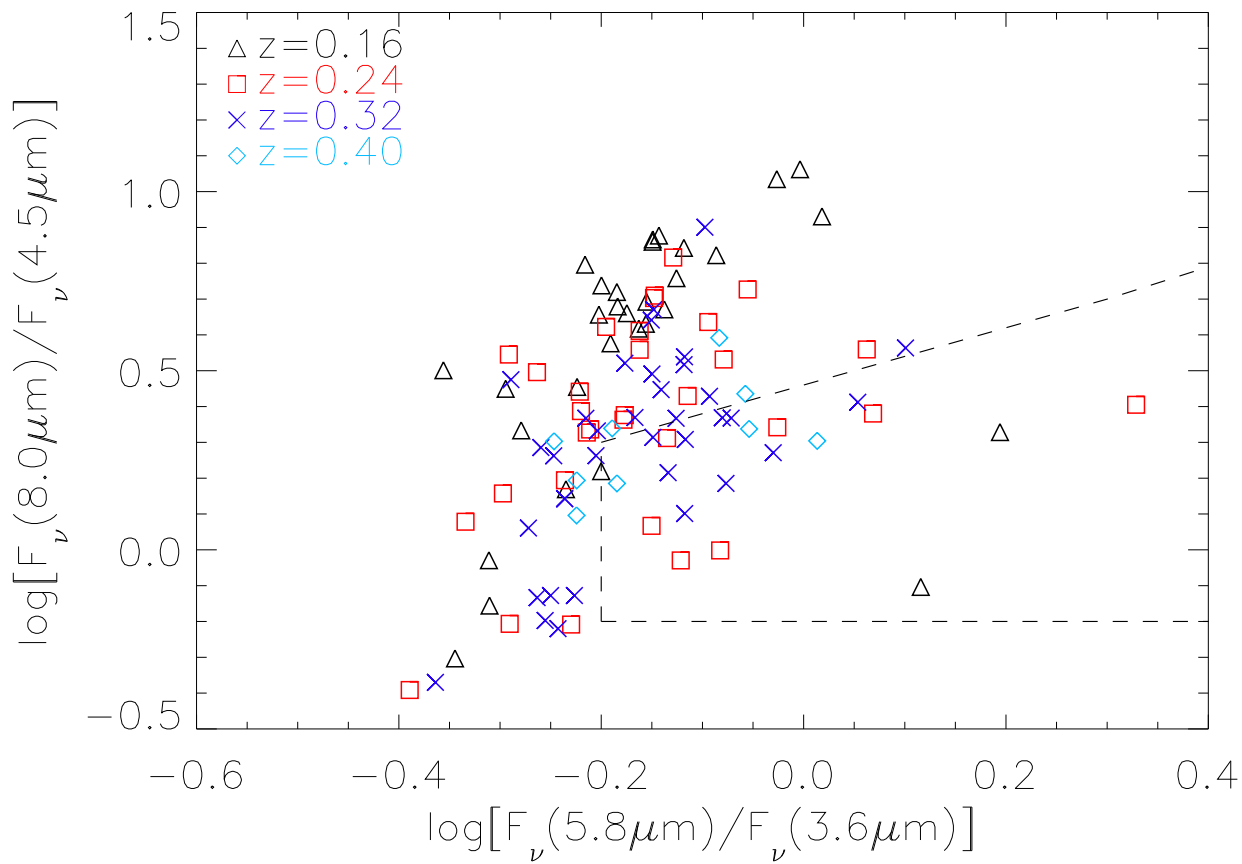}
\caption{An IRAC color-color plot of the WySH sources.  Triangles represent sources at $z~\sim~0.16$, squares represent sources at $z~\sim~0.24$, crosses represent sources at $z~\sim~0.32$ and diamonds represent sources at $z~\sim~0.40$.  The region enclosed by the dashed line shows the empirically determined color criteria used to distinguish possible AGN within the sample \citep{Lacy:2004}.}
\label{agn}
\end{figure}

The empirical method of \citet{Lacy:2004} is used to identify possible AGN within our sample (see Figure~\ref{agn}).  The different ``sequences'' identified in \citet{Lacy:2004} are not immediately apparent in Figure~\ref{agn} since the large Sloan Digital Sky Survey (SDSS) dataset is not also included here as in Figure~1 of \citet{Lacy:2004}.  \citet{Lacy:2004} describe one ``sequence'' to have blue colors in $L_{\nu}(5.8\mu{\rm m})/L_{\nu}(3.6\mu{\rm m})$ and very red colors in $L_{\nu}(8.0\mu{\rm m})/L_{\nu}(4.5\mu{\rm m})$, which they describe as likely to be low-redshift ($z~\lesssim~0.2$) galaxies whose PAH emission bands have not been significantly redshifted out of the 8.0$\mu$m filter.  Within this same sequence but at higher redshifts, the 7.7$\mu$m PAH feature is no longer observed with the 8.0$\mu$m filter, which would give those sources a bluer color in $L_{\nu}(8.0\mu{\rm m})/L_{\nu}(4.5\mu{\rm m})$, and a redder color in $L_{\nu}(5.8\mu{\rm m})/L_{\nu}(3.6\mu{\rm m})$ since the 3.3$\mu$m PAH feature will have redshifted out of the 3.6$\mu$m band.  In Figure~\ref{agn}, it is clear that most of the sources in this sequence as defined by \citet{Lacy:2004} are found at either $z~\sim~0.16$ or $z~\sim~0.24$, however, the sources at $z~\sim~0.32$ and $z~\sim~0.40$ start to span the gap between the two sequences due to the effects of redshift and lie closer to, or within, the enclosed region.  The other ``sequence'' that \citet{Lacy:2004} describe has red colors in both filter pairs; this is where they find SDSS- and radio-selected quasars.  The dashed lines in Figure~\ref{agn} enclose this region, as defined empirically by \citet{Lacy:2004}.  Regarding the WySH sample, the \citet{Lacy:2004} empirical classification of AGN is used to conservatively identify the three right-most sources within the box in Figure~\ref{agn} as AGN.  These three sources are removed from the remainder of this study.  The identities are less clear for the other sources near the AGN/star-forming boundaries, which are indicated by dashed lines.  It is possible that some of these sources are composite sources (i.e., a combination of AGN and star formation within the galaxy).  Other sources, particularly at $z~\sim~0.32$ and $z~\sim~0.40$, may have simply been redshifted into this region based on the previous description of the ``sequences''.  Without spectra, it is impossible to confirm whether or not the sources in question are AGN, star forming or a composite, and these sources near the AGN/SF boundaries are not removed from the sample.  

\section{RESULTS}
\label{anal-res}

\subsection{Dust Attenuation via 24$\mu$m \& FUV}
\begin{figure}
\includegraphics{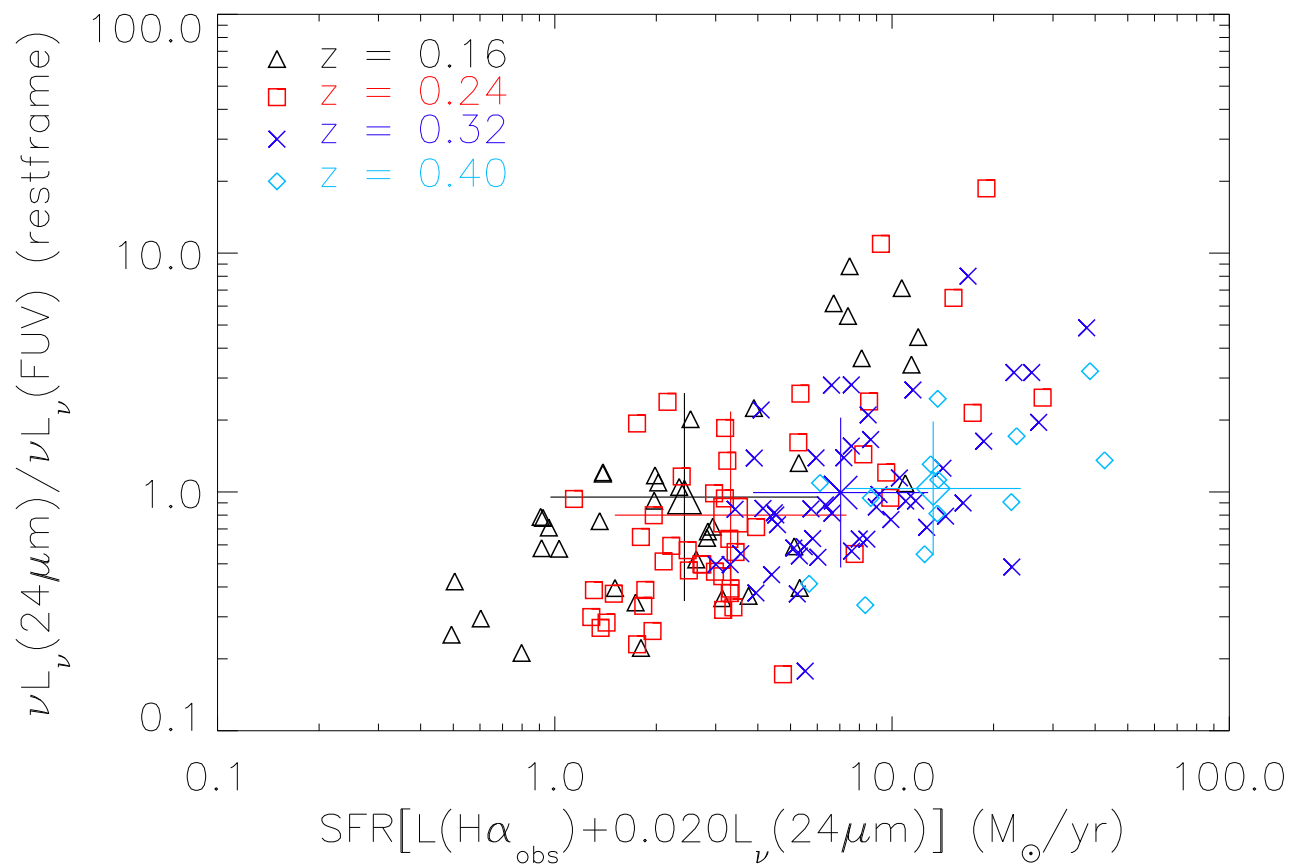}
\caption{A proxy for the traditional measure of dust attenuation, $\nu L_{\nu}(24\mu{\rm m})/\nu L_{\nu}(FUV)$ (in the restframe) plotted vs. the H$\alpha$+24$\mu$m star formation rate from \citet{Kennicutt:2009}.  Triangles represent sources at $z~\sim~0.16$, squares represent sources at $z~\sim~0.24$, crosses represent sources at $z~\sim~0.32$ and diamonds represent sources at $z~\sim~0.40$.  Large symbols represent the average value for each epoch, with error bars indicating the 1$\sigma$ standard deviation.}
\label{24toFUV}
\end{figure}

Very few sources in this study have detections of far-infrared emission, so the traditional $L(TIR)/L(FUV)$ tracer of the attenuation by dust cannot be used, and is replaced here with $\nu L_{\nu}(24\mu{\rm m})/\nu L_{\nu}(FUV)$ as proxy.  The rest-frame 24$\mu$m emission is approximately $10\%$ of the TIR emission on average for normal star forming galaxies, but the dispersion is large.  For normal star-forming galaxies at the epochs studied here, the full range in 24$\mu$m/TIR spans a factor of 2 \citep[][see Figure~16]{Dale:2005}.  Figure \ref{24toFUV} shows the dust attenuation diagnostic, $\nu L_{\nu}(24\mu{\rm m})/\nu L_{\nu}(FUV)$ as a function of the H$\alpha$+24$\mu$m star formation rate from \citet{Kennicutt:2009}:
\begin{equation}
  SFR({\rm M_{\odot} yr^{-1}}) = 7.9 \times 10^{-42}[aL({\rm H}\alpha _{\rm obs})+bL_{\nu}(24\mu{\rm m})]. \\
\label{kennicutt_sfr}
\end{equation}  
where both luminosities are in ${\rm erg~s^{-1}}$, and $L_{\nu}(24\mu{\rm m})$ is expressed as $\nu L_{\nu}$.  The best fit to the data from \citet{Kennicutt:2009} gives $b/a=0.020$, which is based on the integrated fluxes of a large sample of galaxies.  Figure~\ref{24toFUV} follows the trend that galaxies with larger star formation rates tend to be more dusty, as inferred from the 24$\mu$m/FUV ratio \citep{Hopkins:2001a}.  Looking at only the star formation rate portrayed along the x-axis, a trend toward higher star formation rates at higher redshifts is seen, which has been shown by a number of other studies \citep[e.g.,][]{Cowie:2004, LeFloch:2005, Doherty:2006, Dale:2008, Prescott:2009}.  This evolution could partially be due to sample selection whereby only the brighter galaxy subsets are observed at higher redshifts.  The data from all four epochs appear to have roughly the same average attenuation by dust, as indicated by the large symbols in Figure~\ref{24toFUV} (and the information portrayed in Figure~\ref{extinction}).  This result agrees with the recent results of \citet{Buat:2009}.  While the ensemble average is relatively constant from epoch to epoch, each epoch separately exhibits a larger attenuation by dust for higher star formation rates.  Hence, given that the average galaxy star formation rate increases with lookback time, an epoch to epoch comparison at a fixed star formation rate suggests a mild decrease in dust attenuation with redshift.  The argument of a decrease in the amount of attenuation by dust with lookback time is consistent with results from \citet{Buat:2007a} and combined results from \citet{Reddy:2006, Buat:2007b} and \citet{Burgarella:2007}, who have found dust attenuation values decrease from $z \sim 0$ to $z \sim 2$, as discussed previously.

\subsection{Dust Attenuation via H$\alpha$ \& 24$\mu$m}
\begin{figure}
\includegraphics{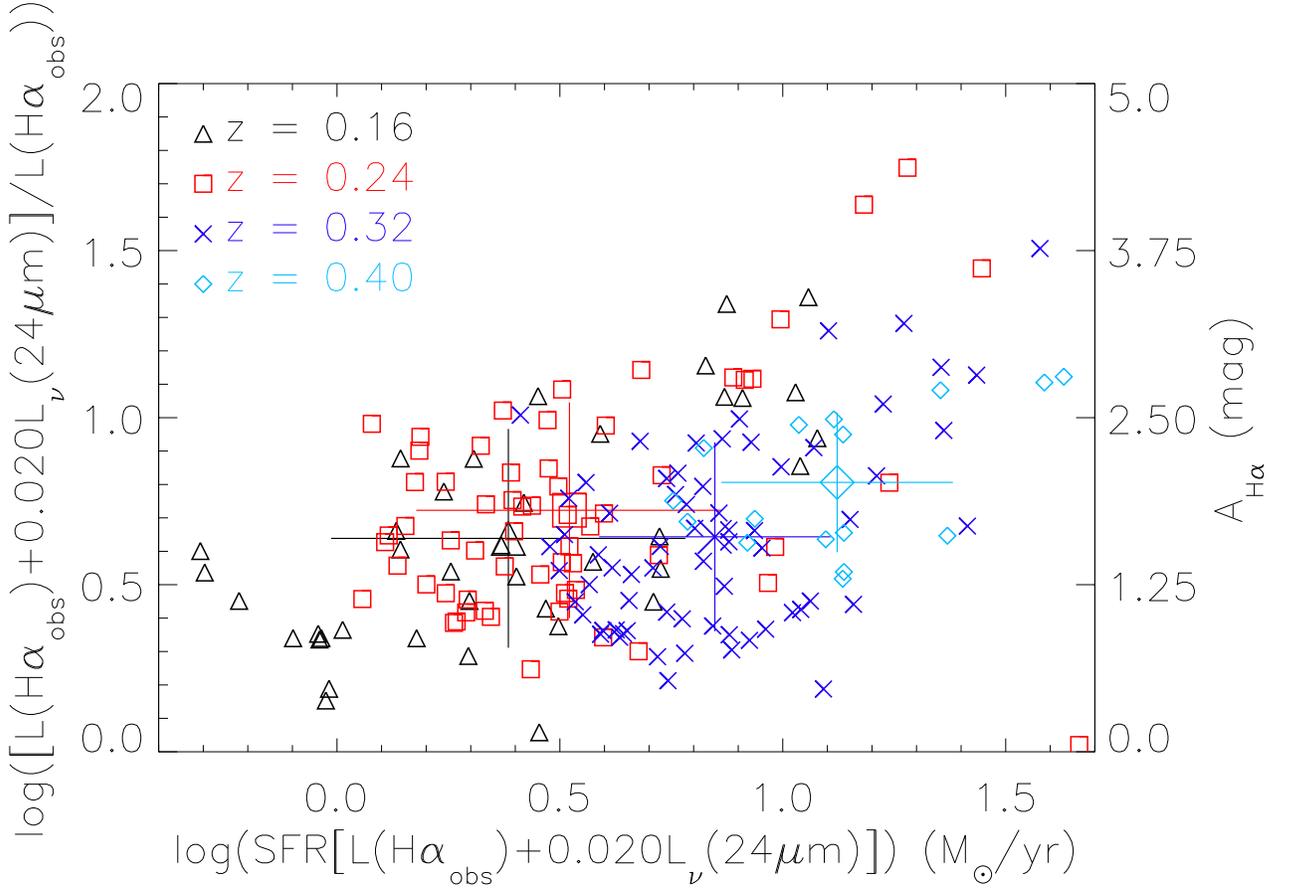}
\caption{The recent diagnostic of dust attenuation, $[aL({\rm H}\alpha_{\rm obs})+bL_{\nu}(24\mu{\rm m})]/L({\rm H}\alpha_{\rm obs})$, where the numerator is taken from \citet{Kennicutt:2009} as a function of the H$\alpha$+24$\mu$m star formation rate from \citet{Kennicutt:2009}, and $b/a=0.020$.  $A_{\rm H\alpha}$ \citep{Kennicutt:2007} is plotted on the right hand axis.  Symbols are the same as in Figure \ref{24toFUV}.}
\label{Ha24toHa}
\end{figure}

The empirical argument for a decrease in dust attenuation is perhaps more readily apparent from the data portrayed in Figure~\ref{Ha24toHa}.  Figure~\ref{Ha24toHa} shows the dust attenuation diagnostic, $[aL({\rm H}\alpha_{\rm obs})+bL_{\nu}(24\mu{\rm m})]/L({\rm H}\alpha_{\rm obs})$ versus the star formation rate indicator of \citet{Kennicutt:2009} (Equation~\ref{kennicutt_sfr}).  Here, again, the expected trend that galaxies with higher star formation rates tend to be more dusty can be seen \citep{Hopkins:2001a}, although, in this case the trend is seen most strongly for a given redshift rather than for the entire dataset.  The right hand axis translates this diagnostic into a value of magnitudes of $A_{{\rm H}\alpha}$ extinction \citep{Kennicutt:2007}.  The large symbols in Figure~\ref{Ha24toHa} represent the average values for each epoch.  As in Figure~\ref{24toFUV}, the average dust attenuation remains relatively constant from epoch to epoch.  Moreover, it is likewise more illuminating to compare the attenuation from epoch to epoch at a fixed star formation rate.  For a given star formation rate, the typical amount of attenuation by dust decreases with increasing redshift, echoing the combined results of \citet{Reddy:2006}, \citet{Buat:2007b}, and \citet{Burgarella:2007}.  While these studies use a UV selection, no comparable studies exist using an H$\alpha$ selection.  For work based on other selections, see, for example, \citet{Buat:2007a}, \citet{Xu:2007}, and \citet{Zheng:2007}.  These comparisons do not change the result that while the ensemble average of dust attenuation does not change as a function of the redshifts sampled here, we expect higher redshift galaxies to show higher values of attenuation since they are typically exhibiting larger star formation rates.  However, we find that higher redshift galaxies show lower values of attenuation at a given luminosity (i.e., star formation rate).  

\subsection{Comparison of Dust Attenuation Diagnostics}
To compare Figures~\ref{24toFUV} and \ref{Ha24toHa}, the correlation coefficients for each diagnostic were computed.  The $\log(\nu L_{\nu}(24\mu{\rm m})/\nu L_{\nu}(FUV))$ diagnostic gives $r=0.545$ and the $\log([aL({\rm H}\alpha_{\rm obs})+bL_{\nu}(24\mu{\rm m})]/L({\rm H}\alpha_{\rm obs}))$ diagnostic gives $r=0.452$.  To investigate the statistical significance of the two diagnostics, the program R\footnote{http://www.r-project.org/} \citep{rstats} was used to analyze the data with a linear model.  Based on the multiple $R^{2}$ statistic, $\sim$0.11\% of the variability in $\log(\nu L_{\nu}(24\mu{\rm m})/\nu L_{\nu}(FUV))$ can be explained by the star formation rate, and $\sim$0.01\% of the variability in $\log([aL({\rm H}\alpha_{\rm obs})+bL_{\nu}(24\mu{\rm m})]/L({\rm H}\alpha_{\rm obs}))$ can be explained by a linear association with $\log(SFR)$.  It can be argued that the latter diagnostic is slightly more correlated with SFR because the $[aL({\rm H}\alpha_{\rm obs})+bL_{\nu}(24\mu{\rm m})]$ parameter appears on both axes, however, based on the statistical analysis described above, this positive correlation is not seen in the data.   

\begin{figure}
\includegraphics{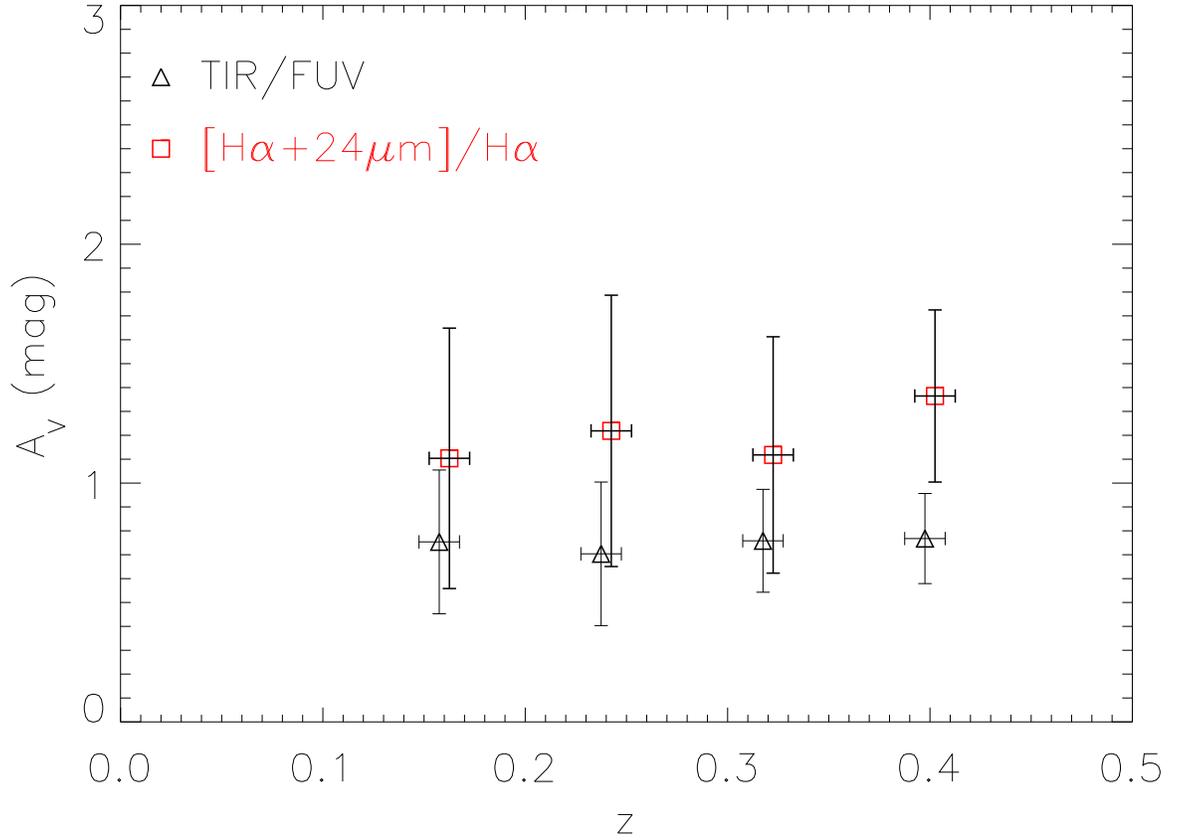}
\caption{Dust attenuation ($A_{V_{cont}}$) in magnitudes v. redshift for both diagnostics.  Triangles with error bars represent the average $A_{V_{cont}}$ at each epoch for $\nu L_{\nu}(24\mu{\rm m})/\nu L_{\nu}(FUV)$.  Squares with error bars represent the average $A_{V_{cont}}$ at each epoch for $[aL({\rm H}\alpha_{\rm obs})+bL_{\nu}(24\mu{\rm m})]/L({\rm H}\alpha_{\rm obs})$.  Error bars are 1$\sigma$ standard deviations of the data.  For clarity, data points have been shifted just to the left and right of each redshift.}
\label{extinction}
\end{figure}

The dust attenuation seems to be higher when looking at the line emission from H$\alpha$ (Fig.~\ref{Ha24toHa}) versus the UV continuum (Fig.~\ref{24toFUV}).  However, it is more informative to look at the amount of attenuation by dust determined from both diagnostics on the same scale ($A_{V_{cont}}$ (continuum) in magnitudes) with respect to redshift (see Figure~\ref{extinction}).  In order to convert $\nu L_{\nu}(24\mu{\rm m})/\nu L_{\nu}(FUV)$ to $A_{V_{cont}}$, $\nu L_{\nu}(24\mu{\rm m})$ was multiplied by 10 to approximate TIR \citep[][Fig. 16]{Dale:2005}.  {\bf (This is a large source of uncertainty, approximately a factor of two.)}  The following prescription from \citet{Buat:2007a} was used to determine $A_{V_{cont}}$ (in magnitudes) from $TIR/FUV$:
\begin{equation}
  A(FUV)=-0.0333\left(\log \frac{L_{TIR}}{L_{FUV}}\right)^{3}+0.3522\left(\log \frac{L_{TIR}}{L_{FUV}}\right)^{2}+1.1960\left(\log \frac{L_{TIR}}{L_{FUV}}\right)+0.4967, \\
\label{afuv}
\end{equation}
\begin{equation}
  A(FUV)=2.678A_{V_{cont}}.\\
\label{afuv-to-av}
\end{equation}
The reddening curve from \citet{Li:2001} converts from $A_{\rm{H}\alpha}$ to $A_{V_{em}}$(emission line) for the $[aL({\rm H}\alpha_{\rm obs})+bL_{\nu}(24\mu{\rm m})]/L({\rm H}\alpha_{\rm obs})$ diagnostic.  The value $R'_{V}=4.05$ is used to convert from $A_{V_{em}}$ to $A_{V_{cont}}$, $A_{V_{em}}\sim1.8A_{V_{cont}}$ \citep{Calzetti:2000, Calzetti:2001}.  As can be seen in Fig.~\ref{extinction}, when the two diagnostics are placed on the same $A_{V_{cont}}$ scale, they agree within the errors.  This consistency holds whether the errors are determined through population statistics or formal measurement uncertainties.  Both diagnostics show no noticeable evolution in the amount of dust attenuation as a function of redshift. 

There are additional challenges to comparing $\nu L_{\nu}(24\mu{\rm m})/\nu L_{\nu}(FUV)$ and $[aL({\rm H}\alpha_{\rm obs})+bL_{\nu}(24\mu{\rm m})]/L({\rm H}\alpha_{\rm obs})$.  The H$\alpha$ and the infrared continuum emission both trace star formation on time scales of $\lesssim10^{7}$yr, while the ultraviolet continuum traces SFRs on time scales of $\sim10^{8}$yr \citep[e.g.,][]{Kennicutt:1998, Meynet:2000, Iglesias-Paramo:2004}.  Thus the $[aL({\rm H}\alpha_{\rm obs})+bL_{\nu}(24\mu{\rm m})]/L({\rm H}\alpha_{\rm obs})$ is naturally a `cleaner' diagnostic of attenuation by dust since all factors involved arise from similar regions, which are physically separated from much of the UV emission, as well as similar time scales \citep[for a counter-argument, see][]{Gordon:2000, Buat:2005, Cortese:2009}.  An additional challenge to interpreting global $\nu L_{\nu}(24\mu{\rm m})/\nu L_{\nu}(FUV)$ measures lies in the average dust temperature of a galaxy.  The 24$\mu$m emission is not only sensitive to the {\it amount} of dust, which directly leads to attenuation, but it is also sensitive to the {\it temperature} of the dust.  In the dust models of \citet{Draine:2007}, the flux at 24$\mu$m largely represents emission from very small grains with effective radii of 15-40\AA.  These intermediate-sized dust grains are responsible for a widely varying portion of the bolometric infrared luminosity depending on the intensity of the interstellar radiation field.  As the intensity of the radiation field increases linearly, the 24$\mu$m emission from the dust grains will increase more rapidly, appearing over-luminous with respect to the linear increase of the radiation field \citep[see][]{Calzetti:2007}.  Naturally, any monochromatic indicator of the dust emission is limited in its ability to effectively encapsulate the emission processes of all dust grains that emit between 1 and 1000$\mu$m, and the 24$\mu$m bandpass is no exception.  

\subsection{Comparison with Dust Attenuation Results at $z~\sim~0$}
\begin{figure}
\includegraphics{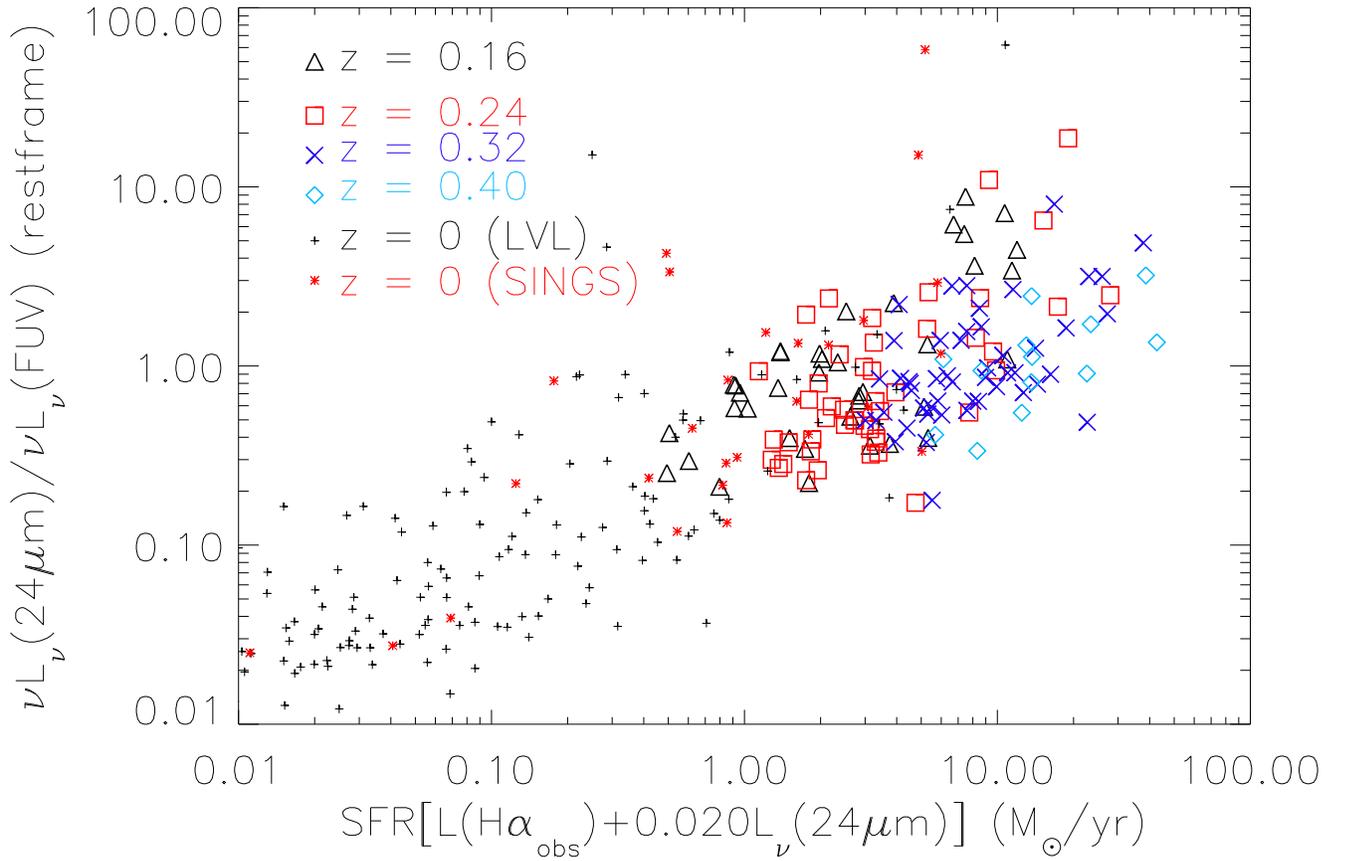}
\caption{Same as Figure~\ref{24toFUV} with $z~\sim~0$ data from the {\it Spitzer} Local Volume Legacy survey added in as small pluses, and data from SINGS added in as small asterisks.}
\label{24toFUVlvl}
\end{figure}

\begin{figure}
\includegraphics{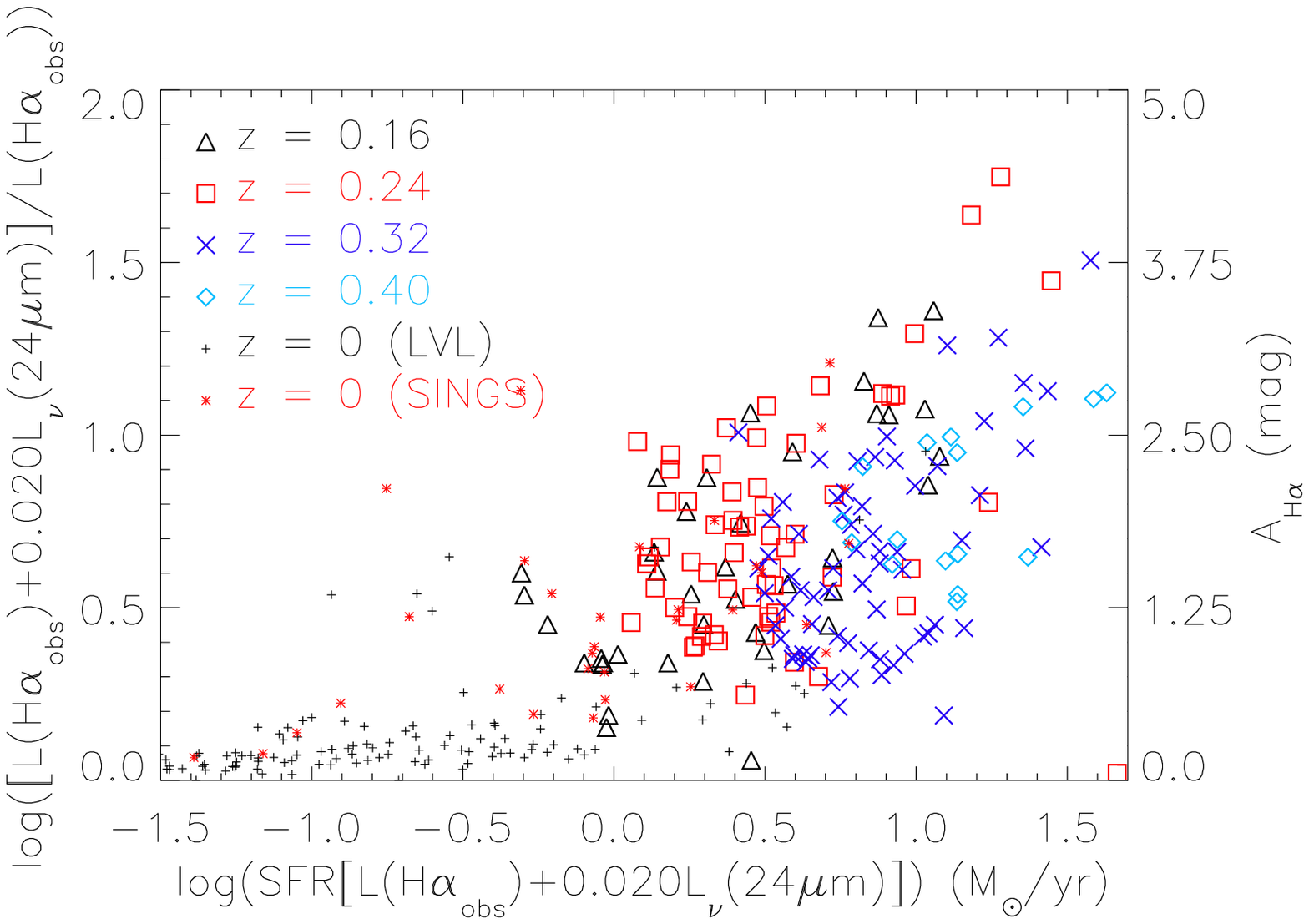}
\caption{Same as Figure~\ref{Ha24toHa} with $z~\sim~0$ data from the {\it Spitzer} Local Volume Legacy survey added in as small pluses, and data from SINGS added in as small asterisks.}
\label{Ha24toHalvl}
\end{figure}

It is also important to compare the WySH dataset to a dataset at $z~\sim~0$, in order to increase the redshift baseline to see how the amount of attenuation by dust evolves from $z~\sim~0$ to $z~\sim~0.40$.  Using the same diagnostics as in Figures~\ref{24toFUV} and \ref{Ha24toHa}, data from the {\it Spitzer} Local Volume Legacy survey and SINGS are added to Figures~\ref{24toFUVlvl} and \ref{Ha24toHalvl} \citep[][Lee et al. 2009, in prep.; for H$\alpha$, 24$\mu$m and UV data, respectively]{Kennicutt:2008, Dale:2009b}.  These samples were chosen because of their multi-wavelength datasets that somewhat mirror the data already being used in this study.  In particular, the {\it Spitzer} Local Volume Legacy survey also contains a larger number of low-luminosity galaxies than would be detected at the redshifts of WySH \citep{Dale:2009b}.  It is clear from Figures~\ref{24toFUVlvl} and \ref{Ha24toHalvl} that the local population of galaxies provides a natural extension to lower SFRs and lower levels of dust attenuation.  Unfortunately, where the WySH and local populations overlap in SFR, there are not enough data to infer any meaningful conclusions.  The emphasis of this comparison is on the fact that the dust attenuation does, in fact, increase with SFR.

\section{SUMMARY}
\label{sum}

The study of the evolution of dust attenuation within galaxies over cosmic time has led to a set of divergent results.  Indeed, the attenuation by dust has been found to decrease from $z~\sim~0$ to $z~\sim~2$ in \citet{Buat:2007a} and \citet{Reddy:2006, Buat:2007b, Burgarella:2007} (combined), to increase from $z~\sim~0$ to $z~\sim~1$ \citep{LeFloch:2005, Schiminovich:2005, Takeuchi:2005} and to remain essentially unchanged from $z~\sim~0.6-0.8$ to the present epoch \citep{Bell:2005, Choi:2006, Xu:2007, Zheng:2007}.  These disparate results could be the effect of different sample selection critiera, with one study focusing on UV-bright galaxies, another on ULIRGS, etc.  The selection criteria for the various studies are summarized in Table~\ref{extinctionstudies}.  There does not appear to be any one single selection criteria that maps to any of the three above-mentioned categories of results (dust attenuation increases with $z$, decreases with $z$, or remains unchanged), so it is difficult to determine whether or not (or to what extent) the selection criteria drive the conclusions made in the previous efforts.  

\begin{deluxetable}{llll}
\rotate
\tabletypesize{\scriptsize}
\tablecaption{Summary of previous studies of dust attenuation as a function of increasing redshift.}
\tablehead{
\colhead{Method\tablenotemark{a}} &
\colhead{Dust Attenuation Amplitude\tablenotemark{b}} &
\colhead{Source Selection} &
\colhead{Reference}
}
\startdata
\cutinhead{dust attenuation decreases}
LD &$<A_{\rm FUV}> \sim3$ mag and decreases &24$\mu$m selection ($z~\lesssim~1$); LIRGs only &\citet{Buat:2007a}\\
&\hspace{12 pt} from $z=0$ to $z=0.7$ & &\\
LF &$A_{\rm FUV}$ decreases from $\sim3.4$ to 1.7 mag &separate FUV \& FIR selections ($z~\sim~0$) &\citet{Buat:2007b}\\
&\hspace{12 pt} from $z=0$ to $z=2$\tablenotemark{c}& &\\
LF &$A_{\rm FUV}$ decreases from $\sim3.4$ to 1.7 mag &UV selection of Lyman break galaxies &\citet{Burgarella:2007}\\
 &\hspace{12 pt} from $z=0$ to $z=2$\tablenotemark{c}  & \hspace{12pt}($z~\sim~1$) &\\
LF &$A_{\rm FUV}$ decreases from $\sim3.4$ to 1.7 mag &BM/BX (optical) criteria similar to Lyman &\citet{Reddy:2006}\\
 &\hspace{12 pt} from $z=0$ to $z=2$\tablenotemark{c} &\hspace{12 pt} break galaxies at $z~\sim~3$ ($z~\sim~2$) &\\
Ind &$A_{\rm H\alpha}$ decreases $\sim 1.5$ mag &24$\mu$m selection with H$\alpha$ \& UV detections &this study\\
&\hspace{12 pt} from $z\sim0.16$ to $z\sim0.40$ &\hspace{12 pt}required &\\
\cutinhead{dust attenuation increases}
LF &co-moving IR energy density &24$\mu$m selection ($z~\lesssim~1$) &\citet{LeFloch:2005}\\
&\hspace{12 pt}$(1+z)^{3.9\pm0.4}$ ($0 \lesssim z \lesssim 1$)\tablenotemark{d} & &\\
LD &$\rho_{1500}=(1+z)^{2.5\pm0.7}$ to $z \sim 1$\tablenotemark{d} &UV selection ($z~\lesssim~3$) &\citet{Schiminovich:2005}\\
LD &$\rho_{\rm dust}/\rho_{\rm FUV}$ increases from $\sim4$ &FUV \& IR luminosity functions from &\citet{Takeuchi:2005}\\ 
 &\hspace{12 pt}to $\sim15$ ($z=0$ to $z\sim1$) & \hspace{12pt} previous works &\\
\cutinhead{dust attenuation same as at $z \sim 0$}
Ind &-- &NIR~+~MIR selection ($z~\sim~0.8$) &\citet{Choi:2006}\\
LD &-- &24$\mu$m selection ($z~\sim~0.7$) &\citet{Bell:2005}\\
Ind &-- &separate NUV \& 24$\mu$m selections of LIRGS &\citet{Xu:2007}\\
 & & \hspace{12 pt} \& ULIRGS ($z~\sim~0.6$) &\\
LD &-- &COMBO-17 photo-z ($0.6 \lesssim z \lesssim 0.8$) &\citet{Zheng:2007}\\
\enddata
\tablenotetext{a}{Results are derived from either a luminosity function (LF), a luminosity density (LD) or a collection of individual galaxies (Ind).}
\tablenotetext{b}{Amplitude of dust attenuation evolution \& what range the evolution is measured over.} 
\tablenotetext{c}{The results from \citet{Buat:2007b, Burgarella:2007} and \citet{Reddy:2006} are combined to determine the results presented in \citet{Burgarella:2007} out to $z \sim 2$.}
\tablenotetext{d}{The results from \citet{LeFloch:2005} and \citet{Schiminovich:2005} are combined to reach the conclusion from \citet{LeFloch:2005} that dust attenuation increases at higher $z$.}
\label{extinctionstudies}
\end{deluxetable}  

In an effort to decipher this problem, we present multi-wavelength results that explore the evolution of the amount of dust attenuation from $z~\sim~0.16$ to $z~\sim~0.40$.  We use H$\alpha$ data from WySH for a uniform initial source selection, and further constrain our sample by selecting according to detections available in IR data from SWIRE and UV data from GALEX.  Using a traditional measure of dust attenuation, $\nu L_{\nu}(24\mu{\rm m})/\nu L_{\nu}(FUV)$, we show that over this range in redshift, there is no noticeable evolution in the ensemble average amount of attenuation by dust of our sources.  Looking over this same range in redshift, with a more recent measure of dust attenuation, $[aL({\rm H}\alpha_{\rm obs})+bL_{\nu}(24\mu{\rm m})]/L({\rm H}\alpha_{\rm obs})$, we also show that there is no noticeable evolution in the ensemble average amount of attenuation by dust.  However, looking epoch by epoch for a given star formation rate, a mild decrease in dust attenuation with lookback time can be seen.  Adding data from a couple of local galaxy samples, spanning both a volume-complete sample ({\it Spitzer} Local Volume Legacy survey) and one that is more geared toward IR-bright galaxies (SINGS) shows no (or very little) evolution in dust attenuation in galaxies from $z~\sim~0$ to $z~\sim~0.40$. Two general trends found in the data are that the dust attenuation increases with star formation rate and that galaxies were probably intrinsically more luminous at times in the past.

Future efforts will incorporate H$\alpha$ data at $z~\sim~0.8$ and $2.2$ from the NEWFIRM Narrowband H$\alpha$ Survey.  We also plan to extract our own flux measurements from the GALEX data.  The NEWFIRM Narrowband H$\alpha$ Survey collaboration is also obtaining spectral data for the H$\alpha$ sources that are identified at $z~\sim~0.8$ and $2.2$ in order to measure the dust attenuation independently using the H$\alpha$/H$\beta$ ratio (Momcheva et al., in preparation).  This will allow for a confirmation of the use of the empirical star formation rate equation determined locally with SINGS data (Eqn.~\ref{kennicutt_sfr}) as a dust attenuation diagnostic tool.


\acknowledgements
This research was funded through the Wyoming NASA Space Grant Consortium (NNG05G165H) and the NSF CAREER program (AST0348990).
The authors would like to acknowledge Alessandro Boselli, Veronique Buat, Denis Burgarella, Robert Kennicutt, Dave Shupe, and Jason Surace for assistance with various aspects of this research and preparation of this manuscript.  The authors would also like to acknowledge the many University of Wyoming and visiting students and teachers who assisted in obtaining the H$\alpha$ data at WIRO.
GALEX data presented in this paper were obtained from the Multimission Archive at the Space Telescope Science Institute (MAST). STScI is operated by the Association of Universities for Research in Astronomy, Inc., under NASA contract NAS5-26555. Support for MAST for non-HST data is provided by the NASA Office of Space Science via grant NAG5-7584 and by other grants and contracts. 
This research has also made use of the NASA/IPAC Infrared Science Archive, which is operated by the Jet Propulsion Laboratory, California Institute of Technology, under contract with the National Aeronautics and Space Administration.  
This work is based in part on observations made with the {\it Spitzer} Space Telescope, which is operated by the Jet Propulsion Laboratory, California Institute of Technology under a contract with NASA.
This research has made use of NASA's Astrophysics Data System Bibliographic Services.

\bibliographystyle{apj}
\bibliography{apj-jour,references}

\begin{thebibliography}{77}
\expandafter\ifx\csname natexlab\endcsname\relax\def\natexlab#1{#1}\fi

\bibitem[{{Bell} {et~al.}(2005)}]{Bell:2005}
{Bell}, E.~F. {et~al.} 2005, \apj, 625, 23

\bibitem[{{Buat} {et~al.}(2007a){Buat}, {Marcillac}, {Burgarella}, {Le Floc'h},
  {Takeuchi}, {Iglesias-Par{\`a}mo}, \& {Xu}}]{Buat:2007a}
{Buat}, V., {Marcillac}, D., {Burgarella}, D., {Le Floc'h}, E., {Takeuchi},
  T.~T., {Iglesias-Par{\`a}mo}, J., \& {Xu}, C.~K. 2007a, \aap, 469, 19

\bibitem[{{Buat} {et~al.}(2009){Buat}, {Takeuchi}, {Burgarella}, {Giovannoli},
  \& {Murata}}]{Buat:2009}
{Buat}, V., {Takeuchi}, T.~T., {Burgarella}, D., {Giovannoli}, E., \& {Murata},
  K.~L. 2009, \aap, 507, 693

\bibitem[{{Buat} {et~al.}(2005)}]{Buat:2005}
{Buat}, V. {et~al.} 2005, \apjl, 619, L51

\bibitem[{{Buat} {et~al.}(2007b)}]{Buat:2007b}
---. 2007b, \apjs, 173, 404

\bibitem[{{Burgarella} {et~al.}(2007){Burgarella}, {Le Floc'h}, {Takeuchi},
  {Huang}, {Buat}, {Rieke}, \& {Tyler}}]{Burgarella:2007}
{Burgarella}, D., {Le Floc'h}, E., {Takeuchi}, T.~T., {Huang}, J.~S., {Buat},
  V., {Rieke}, G.~H., \& {Tyler}, K.~D. 2007, \mnras, 380, 986

\bibitem[{{Calzetti}(2001)}]{Calzetti:2001}
{Calzetti}, D. 2001, \pasp, 113, 1449

\bibitem[{{Calzetti} {et~al.}(2000){Calzetti}, {Armus}, {Bohlin}, {Kinney},
  {Koornneef}, \& {Storchi-Bergmann}}]{Calzetti:2000}
{Calzetti}, D., {Armus}, L., {Bohlin}, R.~C., {Kinney}, A.~L., {Koornneef}, J.,
  \& {Storchi-Bergmann}, T. 2000, \apj, 533, 682

\bibitem[{{Calzetti} \& {Heckman}(1999)}]{Calzetti:1999}
{Calzetti}, D. \& {Heckman}, T.~M. 1999, \apj, 519, 27

\bibitem[{{Calzetti} {et~al.}(1994){Calzetti}, {Kinney}, \&
  {Storchi-Bergmann}}]{Calzetti:1994}
{Calzetti}, D., {Kinney}, A.~L., \& {Storchi-Bergmann}, T. 1994, \apj, 429, 582

\bibitem[{{Calzetti} {et~al.}(2007)}]{Calzetti:2007}
{Calzetti}, D. {et~al.} 2007, \apj, 666, 870

\bibitem[{{Cardiel} {et~al.}(2003){Cardiel}, {Elbaz}, {Schiavon}, {Willmer},
  {Koo}, {Phillips}, \& {Gallego}}]{Cardiel:2003}
{Cardiel}, N., {Elbaz}, D., {Schiavon}, R.~P., {Willmer}, C.~N.~A., {Koo},
  D.~C., {Phillips}, A.~C., \& {Gallego}, J. 2003, \apj, 584, 76

\bibitem[{{Choi} {et~al.}(2006)}]{Choi:2006}
{Choi}, P.~I. {et~al.} 2006, \apj, 637, 227

\bibitem[{{Cortese} \& {Hughes}(2009)}]{Cortese:2009}
{Cortese}, L. \& {Hughes}, T.~M. 2009, \mnras, 400, 1225

\bibitem[{{Cowie} {et~al.}(2004){Cowie}, {Barger}, {Fomalont}, \&
  {Capak}}]{Cowie:2004}
{Cowie}, L.~L., {Barger}, A.~J., {Fomalont}, E.~B., \& {Capak}, P. 2004, \apjl,
  603, L69

\bibitem[{{Dale} \& {Helou}(2002)}]{Dale:2002}
{Dale}, D.~A. \& {Helou}, G. 2002, \apj, 576, 159

\bibitem[{{Dale} {et~al.}(2005)}]{Dale:2005}
{Dale}, D.~A. {et~al.} 2005, \apj, 633, 857

\bibitem[{{Dale} {et~al.}(2007)}]{Dale:2007}
---. 2007, \apj, 655, 863

\bibitem[{{Dale} {et~al.}(2008)}]{Dale:2008}
---. 2008, \aj, 135, 1412

\bibitem[{{Dale} {et~al.}(2009{\natexlab{a}})}]{Dale:2009a}
---. 2009{\natexlab{a}}, \apj, 693, 1821

\bibitem[{{Dale} {et~al.}(2009{\natexlab{b}})}]{Dale:2009b}
---. 2009{\natexlab{b}}, \apj, 703, 517

\bibitem[{{Dale} {et~al.}(2010)}]{Dale:2010}
---. 2010, \apjl, 712, L189

\bibitem[{{Davies} {et~al.}(2009)}]{Davies:2009}
{Davies}, G.~T. {et~al.} 2009, \mnras, 395, L76

\bibitem[{{Doherty} {et~al.}(2006){Doherty}, {Bunker}, {Sharp}, {Dalton},
  {Parry}, \& {Lewis}}]{Doherty:2006}
{Doherty}, M., {Bunker}, A., {Sharp}, R., {Dalton}, G., {Parry}, I., \&
  {Lewis}, I. 2006, \mnras, 370, 331

\bibitem[{{Draine} \& {Li}(2007)}]{Draine:2007}
{Draine}, B.~T. \& {Li}, A. 2007, \apj, 657, 810

\bibitem[{{Engelbracht} {et~al.}(2005){Engelbracht}, {Gordon}, {Rieke},
  {Werner}, {Dale}, \& {Latter}}]{Engelbracht:2005}
{Engelbracht}, C.~W., {Gordon}, K.~D., {Rieke}, G.~H., {Werner}, M.~W., {Dale},
  D.~A., \& {Latter}, W.~B. 2005, \apjl, 628, L29

\bibitem[{{Engelbracht} {et~al.}(2008){Engelbracht}, {Rieke}, {Gordon},
  {Smith}, {Werner}, {Moustakas}, {Willmer}, \& {Vanzi}}]{Engelbracht:2008}
{Engelbracht}, C.~W., {Rieke}, G.~H., {Gordon}, K.~D., {Smith}, J.-D.~T.,
  {Werner}, M.~W., {Moustakas}, J., {Willmer}, C.~N.~A., \& {Vanzi}, L. 2008,
  \apj, 678, 804

\bibitem[{{Gil de Paz} {et~al.}(2007)}]{GildePaz:2007}
{Gil de Paz}, A. {et~al.} 2007, \apjs, 173, 185

\bibitem[{{Gordon} {et~al.}(2000){Gordon}, {Clayton}, {Witt}, \&
  {Misselt}}]{Gordon:2000}
{Gordon}, K.~D., {Clayton}, G.~C., {Witt}, A.~N., \& {Misselt}, K.~A. 2000,
  \apj, 533, 236

\bibitem[{{Heckman} {et~al.}(1998){Heckman}, {Robert}, {Leitherer}, {Garnett},
  \& {van der Rydt}}]{Heckman:1998}
{Heckman}, T.~M., {Robert}, C., {Leitherer}, C., {Garnett}, D.~R., \& {van der
  Rydt}, F. 1998, \apj, 503, 646

\bibitem[{{Hirashita} {et~al.}(2003){Hirashita}, {Buat}, \&
  {Inoue}}]{Hirashita:2003}
{Hirashita}, H., {Buat}, V., \& {Inoue}, A.~K. 2003, \aap, 410, 83

\bibitem[{{Hopkins} {et~al.}(2001{\natexlab{a}}){Hopkins}, {Connolly},
  {Haarsma}, \& {Cram}}]{Hopkins:2001a}
{Hopkins}, A.~M., {Connolly}, A.~J., {Haarsma}, D.~B., \& {Cram}, L.~E.
  2001{\natexlab{a}}, \aj, 122, 288

\bibitem[{{Hopkins} {et~al.}(2001{\natexlab{b}}){Hopkins}, {Irwin}, \&
  {Connolly}}]{Hopkins:2001b}
{Hopkins}, A.~M., {Irwin}, M.~J., \& {Connolly}, A.~J. 2001{\natexlab{b}},
  \apjl, 558, L31

\bibitem[{{Iglesias-P{\'a}ramo} {et~al.}(2004){Iglesias-P{\'a}ramo}, {Boselli},
  {Gavazzi}, \& {Zaccardo}}]{Iglesias-Paramo:2004}
{Iglesias-P{\'a}ramo}, J., {Boselli}, A., {Gavazzi}, G., \& {Zaccardo}, A.
  2004, \aap, 421, 887

\bibitem[{{Jansen} {et~al.}(2001){Jansen}, {Franx}, \&
  {Fabricant}}]{Jansen:2001}
{Jansen}, R.~A., {Franx}, M., \& {Fabricant}, D. 2001, \apj, 551, 825

\bibitem[{{Kennicutt} {et~al.}(2007)}]{Kennicutt:2007}
{Kennicutt}, R.~C. {et~al.} 2007, \apj, 671, 333

\bibitem[{{Kennicutt} {et~al.}(2009)}]{Kennicutt:2009}
---. 2009, \apj, 703, 1672

\bibitem[{{Kennicutt}(1998)}]{Kennicutt:1998}
{Kennicutt}, Jr., R.~C. 1998, \araa, 36, 189

\bibitem[{{Kennicutt} {et~al.}(2008){Kennicutt}, {Lee}, {Funes}, {Sakai}, \&
  {Akiyama}}]{Kennicutt:2008}
{Kennicutt}, Jr., R.~C., {Lee}, J.~C., {Funes}, Jos{\'e}~G., S.~J., {Sakai},
  S., \& {Akiyama}, S. 2008, \apjs, 178, 247

\bibitem[{{Lacy} {et~al.}(2004)}]{Lacy:2004}
{Lacy}, M. {et~al.} 2004, \apjs, 154, 166

\bibitem[{{Lara-L{\'o}pez} {et~al.}(2009){Lara-L{\'o}pez}, {Cepa},
  {Bongiovanni}, {P{\'e}rez Garc{\'{\i}}a}, {Casta{\~n}eda}, {Fern{\'a}ndez
  Lorenzo}, {Povi{\'c}}, \& {S{\'a}nchez-Portal}}]{Lara-Lopez:2009}
{Lara-L{\'o}pez}, M.~A., {Cepa}, J., {Bongiovanni}, A., {P{\'e}rez
  Garc{\'{\i}}a}, A.~M., {Casta{\~n}eda}, H., {Fern{\'a}ndez Lorenzo}, M.,
  {Povi{\'c}}, M., \& {S{\'a}nchez-Portal}, M. 2009, \aap, 505, 529

\bibitem[{{Le Floc'h} {et~al.}(2005)}]{LeFloch:2005}
{Le Floc'h}, E. {et~al.} 2005, \apj, 632, 169

\bibitem[{{Lee} {et~al.}(2009){Lee}, {Ly}, {Moore}, {Salim}, {Dale}, {Finn}, \&
  {Momcheva}}]{Lee:2009a}
{Lee}, J.~C., {Ly}, C., {Moore}, C., {Salim}, S., {Dale}, D., {Finn}, R., \&
  {Momcheva}, I. 2009, \baas, 41, 246

\bibitem[{{Li} \& {Draine}(2001)}]{Li:2001}
{Li}, A. \& {Draine}, B.~T. 2001, \apj, 554, 778

\bibitem[{{Lonsdale} {et~al.}(2003)}]{Lonsdale:2003}
{Lonsdale}, C.~J. {et~al.} 2003, \pasp, 115, 897

\bibitem[{{Ly} {et~al.}(2007)}]{Ly:2007}
{Ly}, C. {et~al.} 2007, \apj, 657, 738

\bibitem[{{Madden} {et~al.}(2006){Madden}, {Galliano}, {Jones}, \&
  {Sauvage}}]{Madden:2006}
{Madden}, S.~C., {Galliano}, F., {Jones}, A.~P., \& {Sauvage}, M. 2006, \aap,
  446, 877

\bibitem[{{Mannucci} {et~al.}(2009)}]{Mannucci:2009}
{Mannucci}, F. {et~al.} 2009, \mnras, 398, 1915

\bibitem[{{Martin} {et~al.}(2005)}]{Martin:2005}
{Martin}, D.~C. {et~al.} 2005, \apjl, 619, L1

\bibitem[{{Martin} {et~al.}(2007)}]{Martin:2007}
---. 2007, \apjs, 173, 415

\bibitem[{{Meurer} {et~al.}(1999){Meurer}, {Heckman}, \&
  {Calzetti}}]{Meurer:1999}
{Meurer}, G.~R., {Heckman}, T.~M., \& {Calzetti}, D. 1999, \apj, 521, 64

\bibitem[{{Meynet} \& {Maeder}(2000)}]{Meynet:2000}
{Meynet}, G. \& {Maeder}, A. 2000, \aap, 361, 101

\bibitem[{{Morrissey} {et~al.}(2007)}]{Morrissey:2007}
{Morrissey}, P. {et~al.} 2007, \apjs, 173, 682

\bibitem[{{O'Halloran} {et~al.}(2006){O'Halloran}, {Satyapal}, \&
  {Dudik}}]{OHalloran:2006}
{O'Halloran}, B., {Satyapal}, S., \& {Dudik}, R.~P. 2006, \apj, 641, 795

\bibitem[{{Papovich} {et~al.}(2006)}]{Papovich:2006}
{Papovich}, C. {et~al.} 2006, \apj, 640, 92

\bibitem[{{Pei} {et~al.}(1999){Pei}, {Fall}, \& {Hauser}}]{Pei:1999}
{Pei}, Y.~C., {Fall}, S.~M., \& {Hauser}, M.~G. 1999, \apj, 522, 604

\bibitem[{{Prescott} {et~al.}(2009){Prescott}, {Baldry}, \&
  {James}}]{Prescott:2009}
{Prescott}, M., {Baldry}, I.~K., \& {James}, P.~A. 2009, \mnras, 397, 90

\bibitem[{{Prescott} {et~al.}(2007)}]{Prescott:2007}
{Prescott}, M.~K.~M. {et~al.} 2007, \apj, 668, 182

\bibitem[{{R Development Core Team}(2009)}]{rstats}
{R Development Core Team}. 2009, {R: A Language and Environment for Statistical
  Computing}, {R Foundation for Statistical Computing}, {Vienna, Austria},
  {{ISBN} 3-900051-07-0}

\bibitem[{{Reddy} {et~al.}(2006){Reddy}, {Steidel}, {Erb}, {Shapley}, \&
  {Pettini}}]{Reddy:2006}
{Reddy}, N.~A., {Steidel}, C.~C., {Erb}, D.~K., {Shapley}, A.~E., \& {Pettini},
  M. 2006, \apj, 653, 1004

\bibitem[{{Reddy} {et~al.}(2008){Reddy}, {Steidel}, {Pettini}, {Adelberger},
  {Shapley}, {Erb}, \& {Dickinson}}]{Reddy:2008}
{Reddy}, N.~A., {Steidel}, C.~C., {Pettini}, M., {Adelberger}, K.~L.,
  {Shapley}, A.~E., {Erb}, D.~K., \& {Dickinson}, M. 2008, \apjs, 175, 48

\bibitem[{{Rosenberg} {et~al.}(2006){Rosenberg}, {Ashby}, {Salzer}, \&
  {Huang}}]{Rosenberg:2006}
{Rosenberg}, J.~L., {Ashby}, M.~L.~N., {Salzer}, J.~J., \& {Huang}, J.-S. 2006,
  \apj, 636, 742

\bibitem[{{Rowan-Robinson} {et~al.}(2008)}]{Rowan-Robinson:2008}
{Rowan-Robinson}, M. {et~al.} 2008, \mnras, 386, 697

\bibitem[{{Salim} {et~al.}(2007)}]{Salim:2007}
{Salim}, S. {et~al.} 2007, \apjs, 173, 267

\bibitem[{{Salim} {et~al.}(2009)}]{Salim:2009}
---. 2009, \apj, 700, 161

\bibitem[{{Schiminovich} {et~al.}(2005)}]{Schiminovich:2005}
{Schiminovich}, D. {et~al.} 2005, \apjl, 619, L47

\bibitem[{{Takeuchi} {et~al.}(2005){Takeuchi}, {Buat}, \&
  {Burgarella}}]{Takeuchi:2005}
{Takeuchi}, T.~T., {Buat}, V., \& {Burgarella}, D. 2005, \aap, 440, L17

\bibitem[{{Teplitz} {et~al.}(2003){Teplitz}, {Collins}, {Gardner}, {Hill}, \&
  {Rhodes}}]{Teplitz:2003}
{Teplitz}, H.~I., {Collins}, N.~R., {Gardner}, J.~P., {Hill}, R.~S., \&
  {Rhodes}, J. 2003, \apj, 589, 704

\bibitem[{{Tresse} {et~al.}(2002){Tresse}, {Maddox}, {Le F{\`e}vre}, \&
  {Cuby}}]{Tresse:2002}
{Tresse}, L., {Maddox}, S.~J., {Le F{\`e}vre}, O., \& {Cuby}, J.-G. 2002,
  \mnras, 337, 369

\bibitem[{{Wang} \& {Heckman}(1996)}]{Wang:1996}
{Wang}, B. \& {Heckman}, T.~M. 1996, \apj, 457, 645

\bibitem[{{Watson} {et~al.}(2009)}]{Watson:2009}
{Watson}, C.~R. {et~al.} 2009, \apj, 696, 2206

\bibitem[{{Wu} {et~al.}(2007){Wu}, {Zhu}, {Cao}, \& {Qin}}]{Wu:2007}
{Wu}, H., {Zhu}, Y.-N., {Cao}, C., \& {Qin}, B. 2007, \apj, 668, 87

\bibitem[{{Wu} {et~al.}(2006){Wu}, {Charmandaris}, {Hao}, {Brandl},
  {Bernard-Salas}, {Spoon}, \& {Houck}}]{Wu:2006}
{Wu}, Y., {Charmandaris}, V., {Hao}, L., {Brandl}, B.~R., {Bernard-Salas}, J.,
  {Spoon}, H.~W.~W., \& {Houck}, J.~R. 2006, \apj, 639, 157

\bibitem[{{Xu} {et~al.}(2003){Xu}, {Lonsdale}, {Shupe}, {Franceschini},
  {Martin}, \& {Schiminovich}}]{Xu:2003}
{Xu}, C.~K., {Lonsdale}, C.~J., {Shupe}, D.~L., {Franceschini}, A., {Martin},
  C., \& {Schiminovich}, D. 2003, \apj, 587, 90

\bibitem[{{Xu} {et~al.}(2007)}]{Xu:2007}
{Xu}, C.~K. {et~al.} 2007, \apjs, 173, 432

\bibitem[{{Zamojski} {et~al.}(2007)}]{Zamojski:2007}
{Zamojski}, M.~A. {et~al.} 2007, \apjs, 172, 468

\bibitem[{{Zheng} {et~al.}(2007){Zheng}, {Dole}, {Bell}, {Le Floc'h}, {Rieke},
  {Rix}, \& {Schiminovich}}]{Zheng:2007}
{Zheng}, X.~Z., {Dole}, H., {Bell}, E.~F., {Le Floc'h}, E., {Rieke}, G.~H.,
  {Rix}, H., \& {Schiminovich}, D. 2007, \apj, 670, 301

\end{thebibliography}

\end{document}